# Quadratic solitons in higher-order topological insulators

## Yaroslav V. Kartashov

*Institute of Spectroscopy, Russian Academy of Sciences, 108840, Troitsk, Moscow, Russia*
*kartashov@isan.troitsk.ru*

I consider higher-order topological insulator (HOTI) created in $\chi^{(2)}$ nonlinear medium and based on two-dimensional generalization of the Su-Schrieffer-Heeger waveguide array, where transition between trivial and topological phases is achieved by shift of the four waveguides in the unit cell towards its center or towards its periphery. Such HOTI can support linear topological corner states that give rise to rich families of quadratic topological solitons bifurcating from linear corner states. The presence of phase mismatch between parametrically interacting fundamental-frequency (FF) and second-harmonic (SH) waves drastically affects the bifurcation scenarios and domains of soliton existence, making the families of corner solitons much richer in comparison with those in HOTIs with cubic nonlinearity. For instance, the internal soliton structure strongly depends on the location of propagation constant in forbidden gaps in spectra of *both* FF and SH waves. Two different types of corner solitons are obtained, where either FF or SH wave dominates in the bifurcation point from linear corner state. Because the waveguides are two-mode for SH wave, its spectrum features two groups of forbidden gaps with corner states of different symmetry appearing in each of them. Such corner states give rise to different families of corner solitons. Stability analysis shows that corner solitons in quadratic HOTI may feature wide stability domains and therefore are observable experimentally. These results illustrate how parametric nonlinear interactions enrich the behavior of topological excitations and allow to control their shapes.

Among a variety of photonic topological insulators theoretically predicted and experimentally realized during the last decade [1-5], higher-order topological insulators (HOTIs) occupy a special niche. While conventional $d$-dimensional topological insulators are known to support $(d-1)$-dimensional topological states at their edges, the most representative feature of the higher-order topological insulator of the $n$-th order $(n > 1)$ created on the basis of a $d$-dimensional lattice is the presence in it in topological phase of the boundary states with effective dimensionality $(d-n)$. Thus, in these intriguing structures the effective dimensionality of the topologically nontrivial boundary modes can be more than one dimension below that of the bulk, see general reviews [6-9] on HOTIs based on periodic and aperiodic lattices discovered in various fields of physics. For instance, second-order two-dimensional HOTI can support effectively zero-dimensional corner states in addition to one-dimensional edge states. Even though corner states are immobile, they are still protected by the topology of the bulk and are not destroyed by the disorder in the underlying lattice if it is not strong enough to close the forbidden spectral gap, where corner states reside. Different types of photonic HOTIs have been suggested and experimentally realised including insulators based on square, honeycomb, and Kagome lattices [10-18], as well as HOTIs based on aperiodic lattices possessing discrete rotational symmetry [19-21]. Among the most representative models of second-order HOTI is the two-dimensional generalization of the Su-Schrieffer-Heeger model [22,23] that has enabled the observation of linear corner states in quadrupole topological phase [13].

Combination of nontrivial topology of the photonic waveguiding structure with strong nonlinear response of optical material introduces desired tunability into properties of topological systems and may lead to qualitatively new propagation regimes of excitations [24,25]. For example, ubiquitous cubic nonlinearity in topological insulators may result in the formation of solitons [26,27] bifurcating from topological boundary modes. In photonic Floquet insulators edge solitons have been observed in [28,29]. Self-action in topological systems may lead to nonlinearity-induced topological phases, with drastic contrast in propagation dynamics of low- and high-power excitations [30-33]. There has been also a surge of interest in realization of nonlinear photonic HOTIs. First example of such structure was experimentally reported in [34], while nonlinear bound states in continuum in HOTIs have been observed in [35]. Since then different types of topological solitons have been theoretically studied and experimentally observed in HOTIs with disclinations [36-38], fractal HOTIs [39,40], and HOTIs with longitudinal parameter modulations [41,42]. Properties of corner solitons in such systems may sometimes resemble those of edge solitons in simpler one-dimensional Su-Schrieffer-Heeger models [43-50], but at the same time, corner solitons in HOTIs may have much richer shapes and distinct stability properties associated with more complex structure of linear spectrum of these systems. Very recently dissipative nonlinear corner modes were also observed in organic polaritonic HOTIs [51,52].

While the majority of works on nonlinear effects in topological systems consider only self-action due to ubiquitous cubic nonlinearity, parametric nonlinear interactions may bring a plethora of new opportunities for control of topological states. They include enhancement of nonlinear interactions of waves of different frequencies in topological systems resulting in efficient higher-harmonic generation in boundary states, both in $\chi^{(2)}$ [53] and $\chi^{(3)}$ [54,55] media. Moreover, parametric interactions in $\chi^{(2)}$ media may support unidirectional quadratic edge solitons in Floquet systems [56] and immobile edge solitons in one-dimensional Su-Schrieffer-Heeger lattices [57-59]. Parametric interactions may be particularly efficient in HOTIs due to strong confinement of light in its corners. Nevertheless, solitons resulting from parametric nonlinear interactions in HOTIs have not been reported yet, to the best of my knowledge.

In this paper I describe the diversity of quadratic topological corner solitons that emerge in HOTI based on two-dimensional Su-Schrieffer-Heeger lattice. The unique feature of this system is that the type of bifurcation from linear corner states in it depends on phase mismatch between two field components that determines the location of propagation constant of soliton with respect to linear spectrum of the system at a given frequency. Thus, two different types of corner solitons are obtained with either fundamental-frequency (FF) or second-harmonic (SH) waves vanishing in the bifurcation point from linear corner state. Using the fact that waveguides are multimode for SH wave, I obtained corner solitons emanating from two different topological gaps for this wave. The analysis shows that several of the obtained corner soliton families can be stable, opening the way for their experimental realization.

I consider the interaction between two co-propagating waves with amplitudes $\psi_1$ (FF wave, frequency $\omega$) and $\psi_2$ (SH wave, frequency $2\omega$) in $\chi^{(2)}$ nonlinear medium with inhomogeneous refractive index landscape that defines two-dimensional Su-Schrieffer-Heeger (SSH) waveguide array. The propagation dynamics of the light fields is described by the following dimensionless system of equations [60-65]:

$$i\frac{\partial \psi_1}{\partial z} = -\frac{1}{2}\left(\frac{\partial^2 \psi_1}{\partial x^2} + \frac{\partial^2 \psi_1}{\partial y^2}\right) - \psi_1^* \psi_2 - \mathcal{R}(x,y)\psi_1,$$
$$i\frac{\partial \psi_2}{\partial z} = -\frac{1}{4}\left(\frac{\partial^2 \psi_2}{\partial x^2} + \frac{\partial^2 \psi_2}{\partial y^2}\right) - \psi_1^2 + \beta\psi_2 - 2\mathcal{R}(x,y)\psi_2. \quad (1)$$

In this system the transverse coordinates $x,y$ are normalized to the characteristic scale $r_0$, the propagation distance $z$ is normalized to the diffraction length $k_1 r_0^2$, where $k_1 = n_1(\omega)\omega/c$ is the wavenumber for FF wave, while $k_2 = n_2(2\omega)2\omega/c$ is the wavenumber for SH wave, $n_1(\omega)$ and $n_2(2\omega)$ are the unperturbed values of the refractive index for corresponding waves, $\beta = (2k_1 - k_2)k_1 r_0^2$ is the scaled phase mismatch between FF and SH waves that for convenience (see below) is included as a linear term into SH wave (instead of nonlinear terms), the field amplitudes are also normalized as $\psi_1 = 2\pi\omega^2 \chi^{(2)} r_0^2 c^{-2} E_1$ and $\psi_2 e^{i\beta z} = 2\pi\omega^2 \chi^{(2)} r_0^2 c^{-2} E_2$, where $E_{1,2}$ are the dimensional fields, $\chi^{(2)}$ is the relevant second-order susceptibility for the employed type I phase matching scheme. Notice that the diffraction coefficient for SH wave is $k_2/k_1 \approx 2$ times lower than that for FW wave. The refractive index distribution in the two-dimensional SSH array is described by the function $\mathcal{R}(x,y) = p\sum_{i,j=1}^{N}\sum_{k=1,4}\mathcal{Q}(x-x_{ik}, y-y_{jk})$, where $N=5$ is the number of unit cells along $x$ and $y$ directions, where each cell contains four identical waveguides $\mathcal{Q} = e^{-(x^2+y^2)/a^2}$ of width $a=0.5$, $p \sim \max(\delta\chi^{(1)})2\pi(\omega r_0/c)^2$ is the depth of array that is determined by the modulation depth of linear susceptibility. Due to different wavenumbers the effective depth of the array for SH wave is $k_2/k_1 \approx 2$ times larger than that for FF wave. Further we set $p=4$ that is consistent with modern technologies of creation of inhomogeneous refractive index landscapes in $\chi^{(2)}$ materials [66,67], including direct fs-laser writing technique that can be applied to $LiNbO_3$ and other typical crystals used in experiments with quadratic solitons.

Topological phase in two-dimensional SSH array is induced by shift of the waveguides in the unit cell [see dashed rectangle in Fig. 1(a) highlighting the unit cell with four waveguides in it]. In uniform array the spacing between all waveguides is identical and equal to $d=3$. Waveguides can be shifted by the distance $\delta/2$ along both $x$ and $y$ axes towards the center of each unit cell ($\delta < 0$) as shown in Fig. 1(a) middle row, or towards the periphery of the unit cell ($\delta > 0$) as shown in Fig. 1(a), bottom row. This introduces dimerization into intracell $d_1 = d+\delta$ and intercell $d_2 = d-\delta$ waveguide spacing leading to corresponding variations of intracell $C_1$ and intercell $C_2$ couplings. When the latter become stronger than intracell coupling at $\delta > 0$, the array enters into topological phase. It is associated with nonzero bulk polarization $\mathbf{P} = (P_x, P_y)$ that is defined for periodic (infinite) array, where for example for FF wave [68]:

$$P_x = \frac{i}{S}\iint_{BZ}\langle u_{\mathbf{k},n}(x,y)|\partial_{k_x}|u_{\mathbf{k},n}(x,y)\rangle dk_x dk_y,$$
$$P_y = \frac{i}{S}\iint_{BZ}\langle u_{\mathbf{k},n}(x,y)|\partial_{k_y}|u_{\mathbf{k},n}(x,y)\rangle dk_x dk_y, \quad (2)$$

where $u_{\mathbf{k},n}(x,y) = u_{\mathbf{k},n}(x+d, y+d)$ is the periodic part of the Bloch function $\psi_1 = u_{\mathbf{k},n} e^{ik_x x + ik_y y + ibz}$ of the array, $S = (2\pi/d)^2$ is the area of the first Brillouin zone, $\mathbf{k} = (k_x, k_y)$ is the Bloch momentum. By analogy with conventional insulators, in HOTIs one can introduce polarization $\mathbf{P} = (1/2, 1/2)$ for the topological gap at $\delta > 0$, where corner states appear, while for array in nontopological regime at $\delta < 0$ one has $\mathbf{P} = (0,0)$ [34,35].

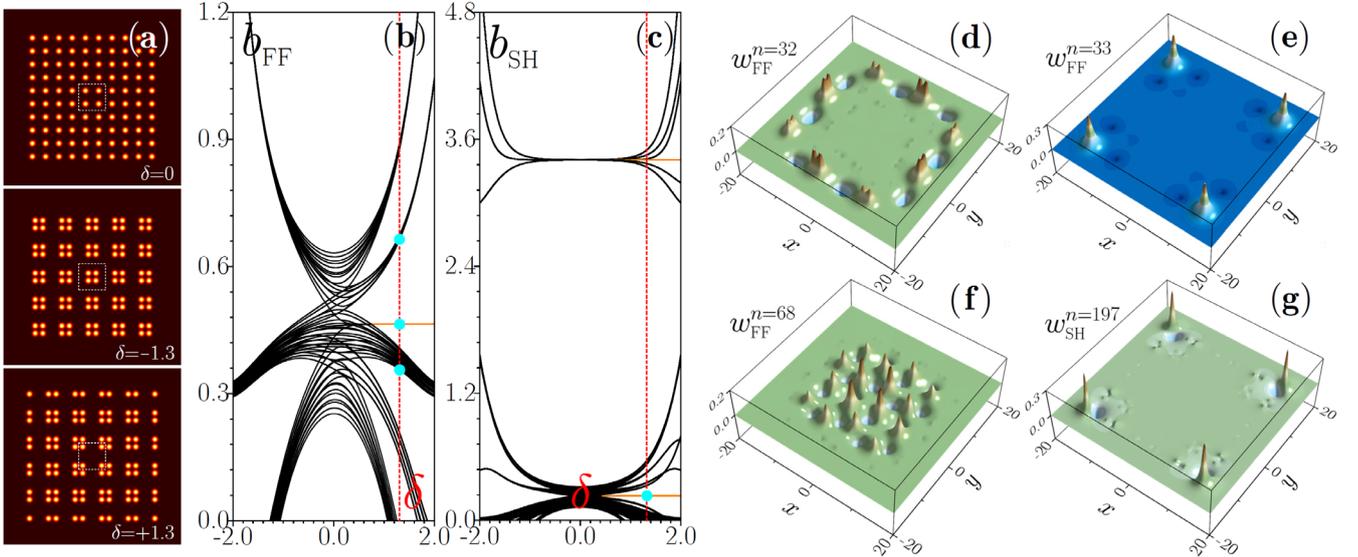

Figure 1. (a) Examples of the waveguide arrays with different waveguide shifts $\delta$ in the unit cell. Dashed square indicates unit cell. Propagation constants of all array eigenmodes versus shift $\delta$ for FF (b) and SH (c) waves. Exemplary profiles of FF edge (d), corner (e), and bulk (f) eigenmodes at $\delta = 1.3$ corresponding to the cyan dots in (b). (g) Profile of SH corner mode at $\delta = 1.3$ corresponding to the cyan dot in (c). In all cases $\beta = 0$.

To illustrate how shift of waveguides in finite array affects linear spectrum of both FF and SH waves, one can omit nonlinear terms in Eq. (1) and search for localized modes in the form $\psi_1 = w_{FF}(x,y)e^{ib_{FF}z}$ for FF wave and $\psi_2 = w_{SH}(x,y)e^{ib_{SH}z}$ for SH wave. Substitution of the fields in such form in Eq. (1) yields two independent linear eigenvalue problems:

$$b_{\text{FF}}w_{\text{FF}} = \frac{1}{2}\left(\frac{d^2w_{\text{FF}}}{dx^2} + \frac{d^2w_{\text{FF}}}{dy^2}\right) + \mathcal{R}w_{\text{FF}},$$
$$b_{\text{SH}}w_{\text{SH}} = \frac{1}{4}\left(\frac{d^2w_{\text{SH}}}{dx^2} + \frac{d^2w_{\text{SH}}}{dy^2}\right) - \beta w_{\text{SH}} + 2\mathcal{R}w_{\text{SH}}, \quad (3)$$

that were solved numerically. Propagation constants of all eigenmodes of the array versus waveguide shift $\delta$ are presented in Fig. 1(b) for FF wave, and in Fig. 1(c) for SH wave. One can observe that increase/decrease of the phase mismatch $\beta$ simply leads to downward/upward shift of the spectrum for SH wave, hence Fig. 1(c) shows results for representative case $\beta=0$. At $\delta \leq 0$ the array is in trivial phase and no localized modes exist in the spectrum, even though there are forbidden gaps. At $\delta > 0$ in FF spectrum top, middle, and bottom groups of propagation constants are associated with bulk modes [an example of such mode is presented in Fig. 1(f) for representative shift $\delta = 1.3$ that I consider further]. The groups of eigenvalues below the top band and above the bottom band are associated with modes localized at the edge of the array [an example is given in Fig. 1(d)]. Finally, orange line corresponds to four practically degenerate propagation constants of topological corner states [Fig. 1(e)]. They differ by phases of spots in different corners. Localization of corner states increases with $\delta$. One can observe in Fig. 1(b) a known property of two-dimensional SSH system, where corner state leaves the band of bulk states for certain nonzero positive $\delta$. Because optical potential for SH wave is two times stronger, while diffraction is weaker, the SH spectrum is characterized by narrow bulk and edge bands [Fig. 1(c)]. Corner states are also shown with orange line. A remarkable property of SH spectrum is that waveguides for this wave become two-mode and due to this two different types of corner modes can be found in spectrum. Lower orange line in Fig. 1(c) corresponds to eight new corner states. Field in such states changes its sign inside the waveguides [see example in Fig. 1(g)], while eight states differ not only by phases of spots in different corners, but also by the orientation of the nodal line inside corner waveguides – in Fig. 1(g) it is perpendicular to the diagonal of the array, while there are also states where this nodal line is parallel to the diagonal.

Topological corner solitons in this system bifurcate from linear corner modes due to parametric interactions. Such solitons can be found in the form $\psi_1 = w_1 e^{ibz}$ and $\psi_2 = w_2 e^{2ibz}$, and they are parameterized by a single propagation constant $b$. To be exponentially localized inside the array and in uniform medium around it, corner soliton should have propagation constant belonging to forbidden gap both in FF and SH components. In other words, localization can take place when $b$ falls into forbidden gap in $b_{\text{FF}}$ spectrum of Fig. 1(b), and when $2b$ falls into forbidden gap in $b_{\text{SH}} - \beta$ spectrum in Fig. 1(c) (here it is taken into account that the latter spectrum shifts linearly with $\beta$). Thus the domains of corner solitons existence in HOTI can be understood from Fig. 2, where on the same $(\beta, b)$ plane I overlaid bands from FF wave spectrum $b_{\text{FF}}$ (gray regions) and bands from transformed $(b_{\text{SH}} - \beta)/2$ SH spectrum (red regions) for representative value of waveguide shift $\delta = 1.3$ corresponding to array in topological phase. Dashed black line in Fig. 2 corresponds to the propagation constant of corner state in FF wave, while dashed blue lines show propagation constants of corner states in SH wave. Looking at this figure, one can see that depending on phase mismatch $\beta$ different situations can be realized. For example, at $\beta = +1.0$ SH bands do not cross the forbidden gap in FF wave, and therefore corner solitons can emerge only from FF corner state. When $\beta > +1.6$ propagation constant of corner mode from top gap in SH wave crosses forbidden gap in FF wave, so potentially bifurcation can occur from corner states either in FF or in SH components. When $\beta < -0.2$ corner state from bottom gap in SH spectrum enters into forbidden gap of FF wave, which means that for this mismatch one can potentially find "second-band" corner solitons with complex internal structure inside waveguides. Finally, there exists a critical value of $\beta \approx -1.74$ at which the upper edge of continuous spectrum for SH wave in Fig. 2 (left outermost red line) becomes higher than upper edge of the topological gap for FF wave. This means that for any $\beta$ value below $-1.74$ any nonlinear corner state with propagation constant $b$ located in topological gap for FF wave automatically falls into continuous spectrum for SH wave and cannot be spatially localized. Notice that in practice the value of phase mismatch $\beta$ can be controlled by heating the sample [65] that allows to consider different regimes of interaction of FF and SH waves. Another viable route is using quasi-phase-matching techniques for control of effective phase mismatch $\beta$. For instance, the results reported here may be relevant for quasi-phase-matched photonic crystals [69].

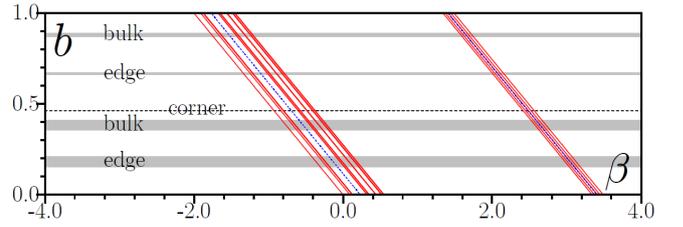

Figure 2. Superimposed allowed bands and forbidden gaps for FF wave [i.e. $b_{\text{FF}}$] and SH wave [i.e. $(b_{\text{SH}} - \beta)/2$] on the $(\beta, b)$ plane at $\delta = 1.3$. Light grey regions correspond to the bands in FF spectrum. For SH spectrum only borders of the bands (red lines) from the first and second groups are plotted. The region below left outermost red line corresponds to continuous part of the SH spectrum. Dashed black and blue lines correspond to eigenvalues of topological corner states in FF and SH spectra, respectively.

The families of topological corner solitons in HOTI were found numerically using Newton relaxation method from the system of equations:

$$-\frac{1}{2}\left(\frac{\partial^2 w_1}{\partial x^2} + \frac{\partial^2 w_1}{\partial y^2}\right) - w_1 w_2 - \mathcal{R}w_1 + bw_1 = 0,$$
$$-\frac{1}{4}\left(\frac{\partial^2 w_2}{\partial x^2} + \frac{\partial^2 w_2}{\partial y^2}\right) - w_1^2 + \beta w_2 - 2\mathcal{R}w_2 + 2bw_2 = 0. \quad (4)$$

The families are represented by the dependencies of total soliton power $U = \iint (w_1^2 + w_2^2)dxdy$ and energy sharing between components $S_{1,2} = U^{-1}\iint w_{1,2}^2 dxdy$ on propagation constant $b$. In all cases, the stability of the obtained solitons was analyzed by adding small (up to 5% in amplitude) noise into initial soliton profiles and propagating them up to very large distance $z = 10^4$ exceeding any experimentally feasible sample length by several orders of magnitude. Solitons were considered stable when their peak amplitude shows only small oscillations due to added noise, and their internal structure does not change at all distances.

The simplest situation is observed when SH bands do not cross the forbidden gap in FF wave spectrum, as it happens, for example, at $\beta = +1.0$ [Fig. 3(a)-3(d)]. In this case corner solitons bifurcate from linear corner mode in FF wave spectrum, whose propagation constant is shown in Fig. 3(a) with vertical black dashed line (in this and similar figures below grey regions illustrate allowed bands for FF wave, while red regions show allowed bands for SH wave). I use logarithmic scale in $U(b)$ plots to clearly show the behavior of soliton near the bifurcation point. In this case bifurcation occurs in the direction of decreasing propagation constants. In the bifurcation point all power is concentrated in FF wave as $S_{1,2}(b)$ plot shows [Fig. 3(d)], while away from bifur-cation point the fraction of power concentrated in SH wave gradually increases. Representative corner soliton profile is shown in Fig. 3(b). It follows from Eq. (4) that while sign of $w_1$ component can be arbitrary, for $w_2$ component it is dictated by the type of solution and $\beta$ value. Thus, for soliton in Fig. 3(b) main spike in $w_2$ component is negative. The soliton gradually expands as it approaches the band of bulk states and becomes delocalized when it enters into this band (I do not show corresponding solutions inside the band, even though they exist). The continuation of this family in the lower gap was obtained too, corresponding solitons gradually expand along the edges (but not into bulk) of the array, as propagation constant $b$ decreases and approaches next band associated with edge states, see example of such soliton in Fig. 3(c). While the branch bifurcating from linear corner state was found stable [as shown by black curve in $U(b)$ dependence], the branch of corner solitons in lower gap is entirely unstable [as shown by the red curve in $U(b)$ dependence].

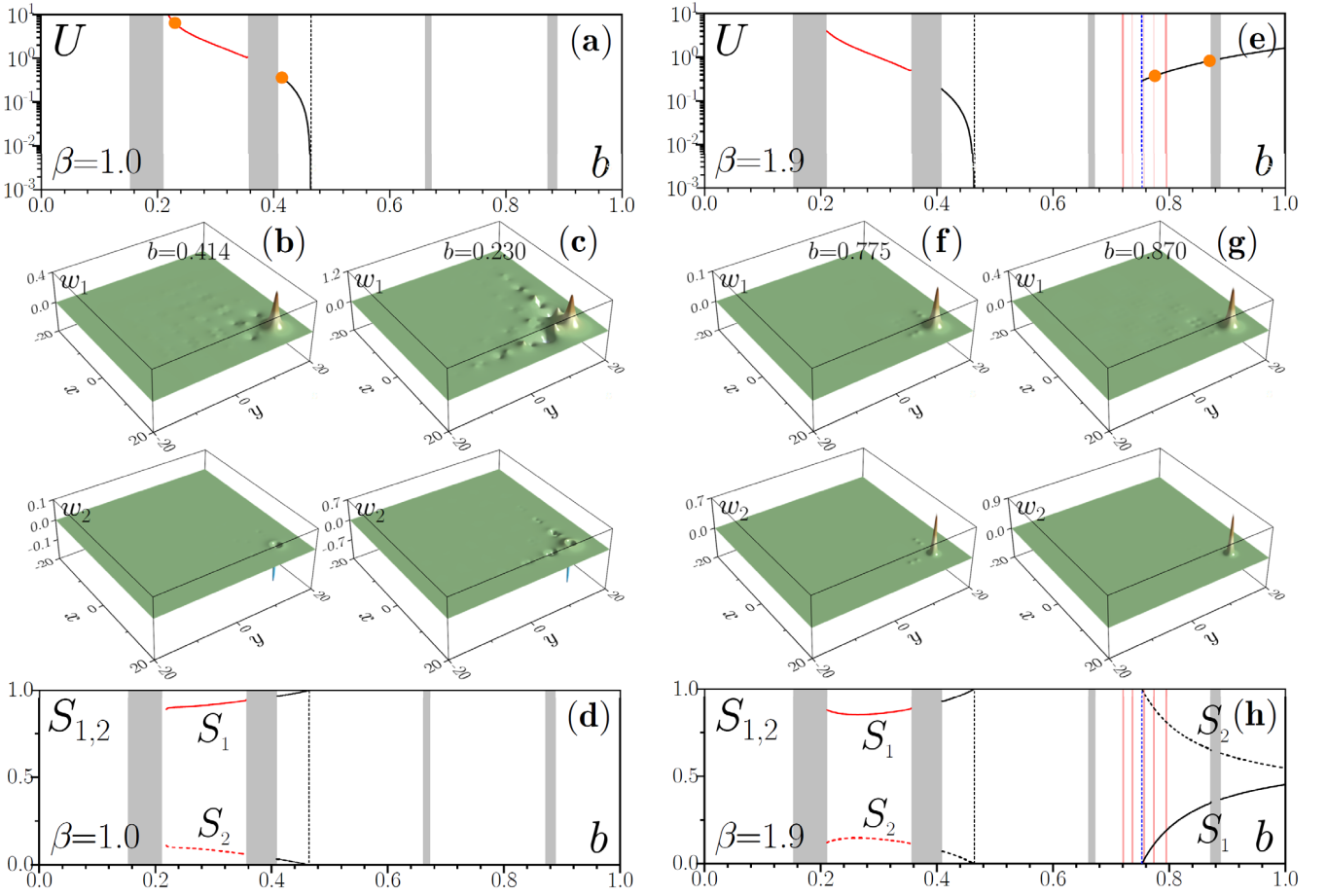

Figure 3. The families $U(b)$ of topological corner solitons in quadratic HOTI at $\beta = 1.0$ (a) and $\beta = 1.9$ (e). Black lines correspond to stable branches, red lines – to unstable branches. Panels (d) and (h) show corresponding dependencies $S_{1,2}(b)$ of energy sharing between FF and SH waves on propagation constant. Grey areas show allowed bands for FF wave, red areas show allowed bands for SH wave. Vertical black and blue dashed lines indicate eigenvalue of linear FF and SH corner modes from which bifurcations occur. Soliton profiles in (b), (c) correspond to orange dots in (a), while profiles in (f), (g) correspond to orange dots in (e). In all cases $\delta = 1.3$.

The picture becomes more intriguing at larger phase mismatch value $\beta = +1.9$ when top group of bands and eigenvalue of corner state from SH wave spectrum "enters" in Fig. 2 into forbidden gap for FF wave. In this case, described in Fig. 3(e)-3(h), the family of corner solitons bifurcating from FF corner state can still be found [see left $U(b)$ branch emerging from black dashed line], but a new family emerging from linear corner state in SH wave appears – see right $U(b)$ branch starting from blue dashed line indicating $(b_{\text{SH}} - \beta)/2$ value for corresponding linear corner state. In this new family all power in the bifurcation point is concentrated in SH wave [Fig. 3(h)], while $S_1$ fraction increases away from bifurcation point. Interestingly, for this new type of solution total power $U$ does not vanish in the bifurcation point. Thus, HOTI with $\chi^{(2)}$ nonlinearity can support two different coexisting types of corner solitons, *both of which* can be stable, as shown with black curves in Fig. 3(e). The representative profiles of corner solitons from new family shown in Fig. 3(f) and 3(g) indicate that $w_2$ component is positive in corner waveguide. The tendency of this state to expand into bulk as its propagation constant approaches the bulk band in FF component is visible in Fig. 3(g), mainly in $w_1$ field.

Further increase of phase mismatch to $\beta = +2.9$ shifts eigenvalue $(b_{\text{SH}} - \beta)/2$ associated with corner state in SH wave below the eigenvalue $b_{\text{FF}}$ of corner state in FF wave spectrum, see Fig. 2. This immediately leads to change of the bifurcation direction of solitons from linear corner states, as illustrated in Fig. 4(a)-4(h). Now bifurcation from linear corner state in FF wave (indicated with black dashed line) occurs in the direction of increasing propagation constants, while bifurcation from linear corner state in SH wave (indicated with blue dashed line) occurs in the direction of decreasing propagation constants [Fig. 4(a)]. The phrase relation between components of such solitons changes as well. While previously in solitons bifurcating from linear FF corner state main peak in $w_2$ component was negative [Fig. 3(b)], for $\beta=+2.9$ it becomes positive [Fig. 4(c)]. In contrast, in soliton bifurcating from SH corner state corner peak in $w_2$ component becomes negative [Fig. 4(b)]. While the family of solitons emanating from FF corner state remains stable even for this phase mismatch, the family of solitons emanating from SH corner state becomes completely unstable (see stable black and unstable red branches in Fig. 4(a)].

Finally, a specific property of this HOTI system is that for depth $p=4$ used here the waveguides for SH wave are two-mode. Due to this the second topological gap appears in linear spectrum of SH wave [Fig. 1(c)] with corner state in it featuring dipole structure inside the waveguides [Fig. 1(g)]. For certain negative value of phase mismatch (for example, $\beta=-0.8$) the eigenvalue of such corner state "enters" into forbidden topological gap for FF wave (Fig. 2). In this case one can also obtain two families of solutions, including usual corner solitons bifurcating from FF corner state, and "second-band" corner solitons bifurcating from linear corner state with complex internal structure (see family emanating from blue dashed line in Fig. 4(e)]. Notice that semi-infinite red region in this figure corresponds to continuum of delocalized states ($b_{\text{SH}}<0$) in spectrum of SH wave, i.e. in this region all solutions become delocalized outside the array. The example of soliton bifurcating from linear FF corner mode in this regime is shown in Fig. 4(f) – corresponding branch of solutions is stable. The example of soliton bifurcating from SH corner mode is shown in Fig. 4(g). One can see that second harmonic in it has dipole structure in all waveguides, while FF component looks like in usual corner soliton, because for this component the waveguides are single-mode. However, this new type of corner solitons is unstable, as shown by the red branch in Fig. 4(e).

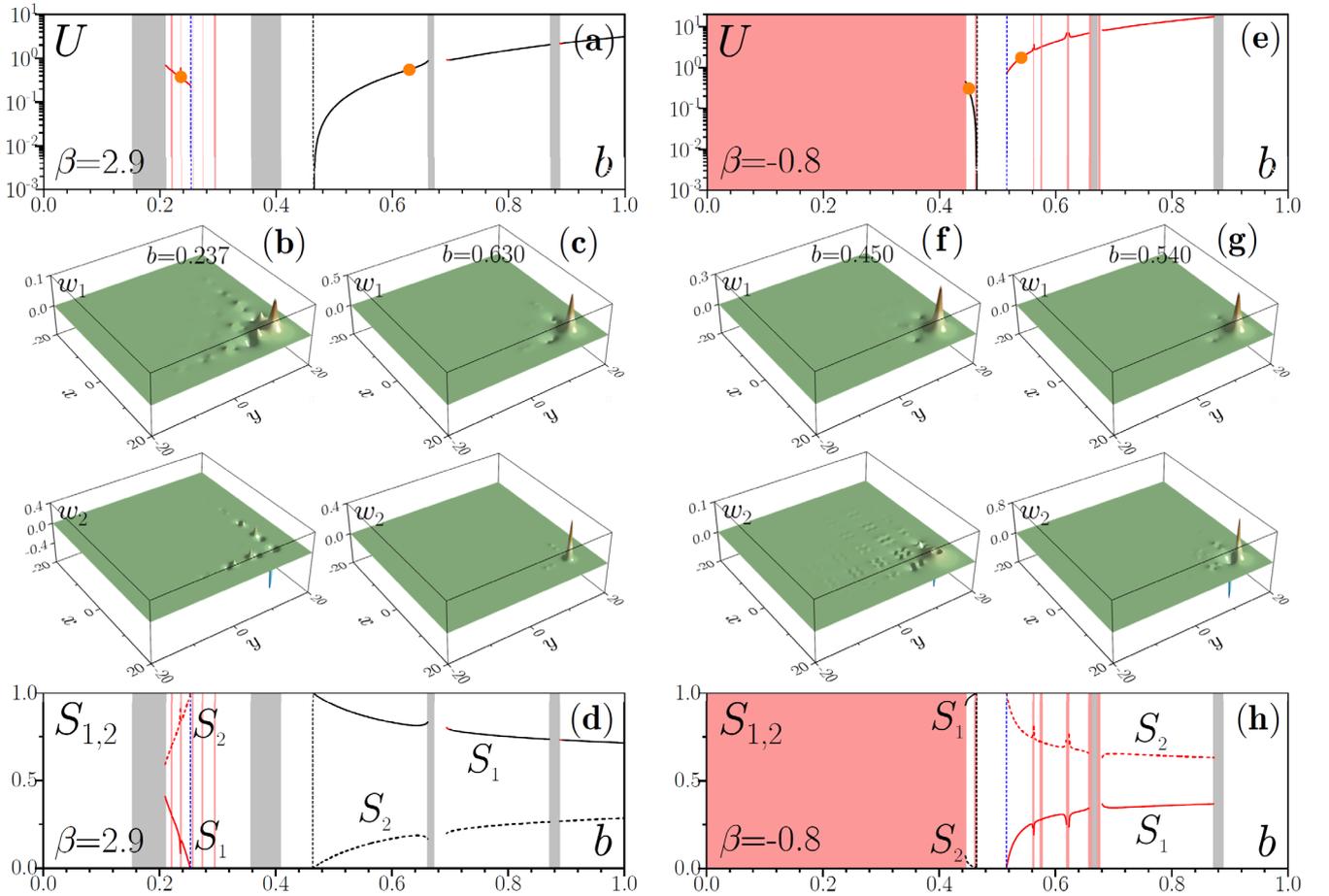

Figure 4. The families $U(b)$ of topological corner solitons in quadratic HOTI at $\beta=2.9$ (a) and $\beta=-0.8$ (e). Black lines correspond to stable branches, red lines – to unstable branches. Panels (d) and (h) show corresponding dependencies $S_{1,2}(b)$ of energy sharing between FF and SH waves on propagation constant. Grey areas show allowed bands for FF wave, red areas show allowed bands for SH wave. Vertical black and blue dashed lines indicate eigenvalue of linear FF and SH corner modes from which bifurcations occur. Soliton profiles in (b), (c) correspond to orange dots in (a), while profiles in (f), (g) correspond to orange dots in (e). In all cases $\delta=1.3$.

Typical examples of propagation dynamics of perturbed quadratic corner solitons are presented in Fig. 5. The evolution was modeled using fast Fourier split-step method, where integration on nonlinear step accounting for interaction between FF and SH waves was implemented using 4$^{\text{th}}$ order Runge-Kutta method. Thus, Fig. 5(a) shows typical instability development for unstable soliton from Fig. 3(c) at $\beta=+1$. The amplitude of this corner soliton shows progressively growing oscillations, it sheds radiation into region outside the array, and eventually, after losing some fraction of input power approaches a new stable corner soliton with lower $U$ (see Visualization 1 showing transformation of the field modulus distribution of the FF wave with distance $z$). Example of completely stable evolution of corner soliton shown in Fig. 3(f) that originates from SH corner state is shown in Fig. 5(c). This state is very well localized in the corner and initial perturbation only causes small amplitude oscillations in the course of propagation (see Visualization 2 showing that FF distribution does not notably change with distance in this case). Finally, instability development for corner soliton from Fig. 4(g) bifurcating from state in second topological gap for SH wave is shown in Fig. 5(c). Notice that even though this soliton is unstable, the instability is weak and leads to slowly growing fast amplitude oscillations of both components (they are visible in Visualization 3 showing FF evolution in this case). The dipole structure of $\psi_2$ field is maintained up to large distances, but eventually this soliton will transform into irregularly breathing corner state.

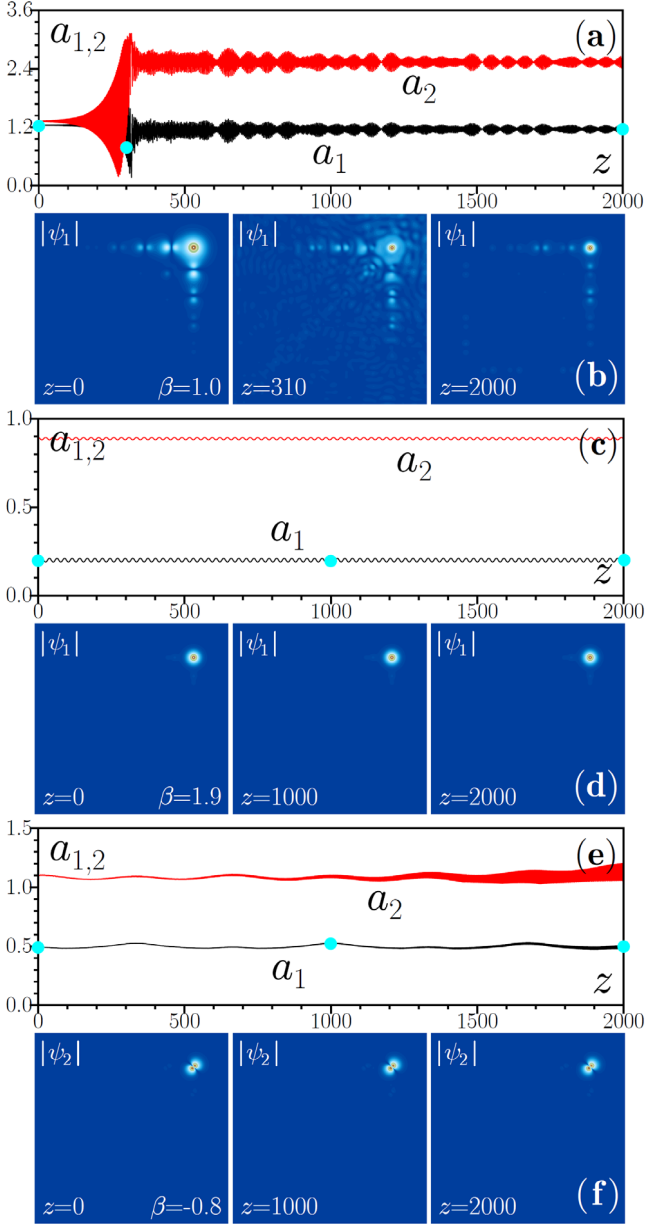

Figure 5. Propagation dynamics of perturbed quadratic corner solitons at $b = 0.230$, $\beta = 1.0$ (see Visualization 1) (a),(b), $b = 0.775$, $\beta = 1.9$ (see Visualization 2) (c),(d), $b = 0.540$, $\beta = -0.8$ (see Visualization 3) (e),(f) in the array with $\delta = 1.3$. In all cases, top panel shows peak amplitudes of FF and SH waves versus distance $z$, while bottom panels show field modulus distribution either in FF or SH wave at different distances corresponding to the cyan dots. Visualizations also show field modulus distributions in $x$- and $y$-cross-sections passing through global intensity maximum of selected component.

Summarizing, I have shown that parametric nonlinear interactions can support a variety of corner solitons in two-dimensional HOTIs. Their internal structure and stability properties are determined by the phase mismatch between FF and SH waves. Both types of corner solitons bifurcating from linear states in FF and SH components can be stable. These results highlight the potential of parametric interactions for realization of new types of topological lasers and control of structure and localization of the wave fields.

**Acknowledgements:** Y.V.K. acknowledges funding by the Russian Science Foundation grant 24-12-00167 and partially by the research project FFUU-2024-0003 of the Institute of Spectroscopy of the Russian Academy of Sciences.